\documentclass[aps,prd,twocolumn,groupedaddress]{revtex4}
\usepackage{graphicx}
\usepackage{dcolumn}
\usepackage{bm}
\begin{document}

\title{Latent heat in the chiral phase transition}

\author{Masaharu Iwasaki}
\email{miwasaki@cc.kochi-u.ac.jp}
\author{Hiyoshi Kiuchi}
\affiliation{Department of Physics, Kochi University, Kochi 780-8520, Japan}

\date{\today}

\begin{abstract}
The chiral phase transition at finite temperature and density is discussed in the framework of the QCD-like gauge field theory. The thermodynamical potential is investigated using a variational approach. Latent heat generated in the first-order phase transition is calculated. It is found that the latent heat is enhanced near the tricritical point and is more than several hundred MeV per quark.
\end{abstract}


\maketitle

\section{INTRODUCTION}

A many-particle system in general possesses two or more different forms of existence called phases. Conversions between these phases are called phase transitions. There are mainly two kinds of phase transitions. One is the first-order phase transition, the other is the second-order one. In the first-order phase transition, as in the conversion from vapor to liquid, the entropy of the system changes discontinuously and a certain amount of heat may be generated as a result. This kind of heat has been called latent heat since named by Joseph Black (1728-1799). On the other hand, in the second-order phase transition, the entropy changes continuously and the latent heat is not given off.

The behavior of the phase transition in quark matter is of fundamental interest \cite{W00}. It is believed that at sufficiently high temperature and/or density the system lies in the chirally symmetric phase. Quark-gluon plasma (QGP) is one of the most important possibilities in this phase. Recently there has been great interest in generating QGP in the experiments. It is expected that QGP may be realized in high energy heavy-ion collisions at CERN Large Hadron Collider(LHC) and at the BNL Relativistic Heavy Ion Collider(RHIC). When the temperature of QGP falls and reaches the critical temperature $(T_c)$, the chiral symmetry in QCD is dynamically broken and we have a phase transition from the chirally symmetric phase into the chiral symmetry broken phase.

The chiral phase transition at finite temperature and/or finite density has been studied by many authors \cite{AY89}-\cite{KMT01}. According to them, it was shown that the transition is of second-order at low density and of first-order at high density. The critical density between them is called a tricritical point \cite{SRS98}-\cite{HJSSV98}. This nature of the phase transition was shown by the NJL model \cite{AY89} and then confirmed also by the QCD-like theory \cite{BCDGP90}-\cite{KMT01}. If the phase transition is of first-order, a latent heat must be generated across the critical temperature as mentioned above. Our main goal here is to evaluate the latent heat in the chiral phase transition at finite temperature and at finite chemical potential with the use of the QCD-like theory.

So far much efforts have been made to study about signals of generation of QGP \cite{MS86}. The generated latent heat may give a signal of QGP. For example, the shock wave can be expected to be produced by the generated latent heat and may be available as a signal. The shock wave is supposed to be given off at the phase transition between the two phases. This process is due to the acceleration of particles by the generated latent heat. Therefore, knowledge about the latent heat is an important prerequisite for analysis of the data in such a high energy collision experiment that is expected to generate QGP.

The outline of the paper is as follows. In the next section, we derive the partition function of the quark matter based on the QCD-like gauge field theory in Landau gauge. In sect.‡V, we perform a mean field approximation and derive the Schwinger-Dyson equation for the effective mass. Then (sect.) using a variational procedure, we solve numerically the Schwinger-Dyson equation with a trial mass function and determine the critical point. In sect.‡X, we compute the latent heat numerically. Sect.‡Y is devoted to summary and discussions.

\section{partition function}

We derive here the effective partition function of the quark matter in the framework of the QCD-like gauge field theory. In this paper, we take $N_c=3$ as the number of color and $N_f=3$ as that of flavor. Let us start with the exact QCD partition function in the path-integral formalism \cite{K89,BL94}:
\begin{eqnarray}
&&Z({\rm T},\mu)={\rm Tr~e}^{-\beta \mathcal{H}} \nonumber\\
&&=N\int D\overline{\Psi} D\Psi DA\exp\left[\int^{\beta}_{0} d\tau\int d^{3}x \mathcal{L}_{\rm QCD}\right],
\end{eqnarray}
where $\beta=T^{-1}$, $\tau=it$ and $\mathcal{H}$ stands for the Hamiltonian of the system. When the chiral symmetry is exact, the QCD Lagrangian $\mathcal{L}_{QCD}$ is given by
\begin{equation}
\mathcal{L}_{QCD}=\mathcal{L}_{q}+\mathcal{L}_{f},
\end{equation}
where Lagrangians of the quarks and gluons are given by
\begin{eqnarray}
\mathcal{L}_{q}&\equiv&\overline{\Psi}(i\gamma^{\mu}\partial_{\mu}+\mu\gamma^{0})\Psi-g\overline{\Psi}\gamma^{\mu}\frac{\lambda^{a}}{2}\Psi A^{a}_{\mu},\\
\mathcal{L}_{f}&\equiv&-\frac{1}{4}F^{\mu\nu}F_{\mu\nu}-\frac{1}{2\alpha}(\partial_{\mu}A^{\mu})^{2} \nonumber\\
&=&\frac{1}{2}A^{\mu a}D_{\mu\nu}^{-1}A^{\nu a}+{\rm nonlinear~terms},
\end{eqnarray}
respectively. Here $\mu$, $g$, $\lambda^{a}$ and $\alpha$ denote chemical potential, the coupling constant, color SU(3) matrices and gauge parameter, respectively. The function $D^{-1}_{\mu\nu}=g_{\mu\nu}\partial^{\eta}\partial_{\eta}-(1-\alpha^{-1})\partial_{\mu}\partial_{\nu}$ is the inverse tree-level gluon propagater.

Next, we neglect the nonlinear terms and the coupling constant is replaced by the running coupling one, which is called QCD-like theory \cite{H83}-\cite{BCDDG88}. Then we can integrate out the gluon field $A$. Moreover to investigate the antiquark-quark pairing correlation, we make use of the Fierz rearrangement to $\mathcal{L}_{QCD}$. If the most attractive term is left, the Lagrangian is expressed by
\begin{eqnarray}
&&{\mathcal L}_{\rm QCD}=\overline{\Psi}(x)G^{-1}(x)\Psi(x) \nonumber\\
&&+\frac{{g'}^2}{2}\int d^{4}y\{\overline{\Psi}(x)\Psi(y)\}D_{F}(x-y)\{\overline{\Psi}(y)\Psi(x)\},
\end{eqnarray}
where ${\rm g'}^{2}\equiv (3+\alpha)g^{2}/27$ and the inverse quark propagator $G^{-1}(x)=(i\gamma^{\mu}\partial_{\mu}+\mu\gamma^{0})$ is introduced. We hereafter fix the gauge parameter as $\alpha=0$ (Landau gauge), which is the usual choice in the QCD-like theory.

Let us now introduce an auxiliary bilocal scalar field $\varphi(x,y)$ representing a wave function of an antiquark-quark pair. In order to use the Stratonovich-Haverd transformation, we take the following identity:
\begin{eqnarray}
1&=&C\int D\varphi^{*}D\varphi \exp \left[-\int d^{4}xd^{4}y\{ \varphi(x,y)-\overline{\Psi}(x)\Psi(y) \}\right. \nonumber\\
&&\left.\times {g'}^{2}D_{F}(x-y)\{ {\varphi}^{*}(x,y)-\overline{\Psi}(y){\Psi}(x)\}\right],
\end{eqnarray}
where $C$ is a normalization constant. Substituting this identity into Eq.(1), the partition function can be rewritten as
\begin{eqnarray}
Z=N'\int D\overline{\Psi}D\Psi D\varphi^{*}D\varphi\exp[-S'],
\end{eqnarray}
where the corresponding action is given by
\begin{eqnarray} 
S'&=&\int d^{4}xd^{4}y\left[ \frac{{g'}^2}{2}|\varphi(x,y)|^{2}D_{F}(x-y)\right. \nonumber\\
&&-\overline{\Psi}(x)\{ \delta(x-y)G^{-1}(y)+\frac{{g'}^2}{2}\varphi^{*}(x,y)D_{F}(x-y)\nonumber\\
&&\left.+\frac{{g'}^2}{2}D_{F}(y-x)\varphi(y,x)\}\Psi(y)\right].
\end{eqnarray}

Here we assume that the auxiliary field is dependent only on the relative coordinates: $\varphi(x,y)=\varphi(x-y)$. This means that the center of mass of the quark-antiquark pair is stationary, because we concentrate on the ground state of the system. Let us define the pairing field $\Delta(x-y)$:
\begin{eqnarray}
\Delta(x-y)\equiv{g'}^{2}D_{F}(x-y)\varphi(y,x).
\end{eqnarray}
Using Fourier transformation, the modified gluon propagator $D_{F}(x-y)$ is expressed as
\begin{eqnarray}
D_{F}(x-y)\equiv\beta^{-1}{\sum_{n}}'\int\frac{d^{3}p}{(2\pi)^{3}}\frac{1}{-p^{2}}{\rm e}^{-ip(x-y)},
\end{eqnarray}
where $p\equiv(i\omega_n,{\bf p}),~x\equiv(-i\tau,{\bf x})$ and $\sum'_{n}$ denotes the boson Matsubara frequency sum over $\omega_{n}=2n\pi\beta^{-1}$ with an integer $n$. In the same way, the auxiliary field is expanded as
\begin{eqnarray}
\varphi(x-y)=\beta^{-1}\sum_{n}\int\frac{d^{3}p}{(2\pi)^{3}}\varphi(p)e^{-ip(y-x)},
\end{eqnarray}
where $\sum_{n}$ denotes the fermion Matsubara frequency sum over $\omega_{n}=2(n+1)\pi\beta^{-1}$ with an integer $n$. Then we can rewrite the pairing field as
\begin{eqnarray}
\Delta(x-y)=\beta^{-1}\sum_{n}\int\frac{d^{3}p}{(2\pi)^{3}}\Delta(p)e^{-ip(y-x)},
\end{eqnarray}
where the Fourier component in the right hand side is defined by
\begin{eqnarray}
\Delta(p)\equiv\beta^{-1}\sum_{n}\int\frac{d^{3}q}{(2\pi)^{3}}\frac{{g'}^2}{-(p-q)^{2}}\varphi(q).
\end{eqnarray}

Using the $\Delta(p)$, we can transform the partition function. After integrating out the quark fields $ \overline{\Psi}$ and $\Psi$, we obtain
\begin{eqnarray}
Z&=N'\int D\varphi^{*}D\varphi\exp[-S_{\rm eff}],
\end{eqnarray}
where the effective action $S_{\rm eff}$ is given as
\begin{eqnarray}
S_{\rm eff}&=&-2\xi V\sum_{n}\int\frac{d^{3}p}{(2\pi)^{3}}\log(\Delta^{2}(p)-\overline{p}^{2}) \nonumber\\
&&+\frac{1}{2}V\sum_{n}\int\frac{d^{3}p}{(2\pi)^{3}}\varphi(p)\Delta(p),
\end{eqnarray}
where $\overline{p}\equiv(i\omega_{n}+\mu,{\bf p})$ and $\xi\equiv N_{c}\times N_{f}=3\times3=9$. Thus our effective action is represented by only the $\Delta(p)$. It is interpreted as an effective mass for the quark so that we call it the effective mass function. The effective action obtained above will play a fundamental role in the next section.

\section{MEAN FIELD APPROXIMATION}

In this section, we will approximate the partition function obtained above. To this end, we make use of the stationary (WKB) approximation for the path integral \cite{K89,BL94}. If the extremum of the integral is realized by $\varphi^{0}$, it must satisfy the following stationary condition:
\begin{equation}
\left.\frac{\delta S_{\rm eff}}{\delta \varphi(p)}\right|_{\varphi=\varphi^{(0)}}=0.
\end{equation}
The $\varphi^{(0)}$ is a mean auxiliary field and has physical meaning that $\varphi^{(0)}=<\overline{\Psi}(x)\Psi(y)>$. The superscript ${0}$ is omitted hereafter. The stationary condition is transformed into the following equation,
\begin{eqnarray}
\Delta(p)=3C_{2}\beta^{-1}\sum_{n}\int\frac{d^{3}q}{(2\pi)^{3}}\frac{{\rm g}^{2}}{-(p-q)^{2}}\frac{\Delta(q)}{\Delta^{2}(q)-\overline{q}^{2}},
\end{eqnarray}
where $C_{2}=(N_c^2-1)/(2N_c)=4/3$ is the quadratic Casimir operator for the color SU(3) group. This is nothing but the Schwinger-Dyson equation for an effective quark mass $\Delta(p)$ in the ladder approximation, which is the same equation as that obtained by the previous authors \cite{KMT98}. Of course we may take a variation of $Delta(p)$ instead of $\varphi(p)$. The stationary condition with respect to $\Delta(p)$ leads to 
\begin{eqnarray}
\varphi(p)=4\xi\frac{\Delta(p)}{\Delta^{2}(p)-\overline{p}^{2}}.
\end{eqnarray}
Substituting this equation into Eq.(13), we get the same Schwinger-Dyson equation.

If the effective quark mass function is replaced by the mean value, we can obtain the partition function at once. Moreover from the partition function, the thermodynamical potential $\Omega$ is given by 
\begin{eqnarray}
\Omega(T,\mu)&\equiv& \frac{T}{V}S^{(0)}_{\rm eff} \nonumber\\
&=&2{\xi}T\sum_{n}\int\frac{d^{3}p}{(2\pi)^{3}}\Bigl[-\log(\Delta^{2}(p)-\overline{p}^{2}) \nonumber\\
&&+\frac{\Delta^{2}(p)}{\Delta^{2}(p)-\overline{p}^{2}}\Bigr].
\end{eqnarray}
The thermodynamical potential $\Omega(T,\mu)$ will be a key function in later discussions on the chiral phase transition and the latent heat.

\section{PHASE DIAGRAM}

We are now in a position to solve the Schwinger-Dyson equation (S-D eq.) numerically and determine the phase diagram of the quark matter. Let us now adopt the modified running coupling of the one-loop order $(\overline{g})$ \cite{H91}. It is obtained through the study using the operator product expansion and renormalization-group equation in QCD. The explicit expression is given by
\begin{eqnarray}
{g}^{2}\longrightarrow\overline{g}^{2}(p)=\frac{2\pi^{2}a}{\log[(-p^{2}+p_{R}^{2})/\Lambda^{2}_{QCD}]},
\end{eqnarray}
where $a=6C_2/(11-2N_f/3)=8/9$ and the parameter $p_{R}$ is introduced in order to regulate the infra-red divergence. Here we have assumed that the running coupling is independent of chemical potential as done in Ref.\cite{KMT98}. It should be noted that the asymptotic freedom in the deep Euclidean region is satisfied by this running coupling.

Instead of solving the S-D equation exactly, we take a variational procedure according to Ref.\cite{KMT98}: we introduce the following trial mass function $(\Delta(p))$ obtained through the study of $T=\mu=0$ case:
\begin{eqnarray}
\Delta(p)=\frac{\sigma}{-p^{2}+p^{2}_{R}}\bigl[ \log {(-p^{2}+p_{R}^{2})/\Lambda^{2}_{QCD}}\bigr]^{\frac{a}{2}-1},
\end{eqnarray}
where $\sigma$ is an order parameter of the chiral symmetry. Then the S-D equation with the running coupling is lead to
\begin{eqnarray}
\Delta(q)=3C_{2}\beta^{-1}\sum_{n}\int\frac{d^{3}p}{(2\pi)^{3}}\frac{\overline{g}^{2}(q,p)}{-(q-p)^{2}}\frac{\Delta(p)}{\Delta^{2}(p)-\overline{p}^{2}}.
\end{eqnarray}
Since $-(q-p)^{2}$ in the denominator is a gluon momentum, the most natural form of the running coupling would be $\overline{g}^{2} (-(q-p)^{2})$. However, the momentum dependence would bring about many difficulties from its angle dependence in actual numerical calculations. Assuming that the effect of the angle dependent part $-2qp{\rm cos}\theta$ in $-(q-p)^{2}$ is negligible on average, we approximate the running coupling \cite{H91}:
\begin{eqnarray}
\overline{g}^{2}(q,p)\simeq\overline{g}^{2}(-(q^{2}+p^{2})).
\end{eqnarray}
Since we use the variational procedure with respect to $\sigma$ instead of the self-consistent iteration method, it is sufficient to consider the equation with the lowest value of $-q^2$: $-q^2=\pi^{2}T^{2}$. Thus the S-D equation we ought to solve is
\begin{eqnarray}
&&\frac{1}{\pi^{2}T^{2}+p_{R}^{2}}\bigl[\log\{(\pi^{2}T^{2}+p_{R}^{2})/\Lambda^{2}_{QCD}\}\bigr]^{\frac{-5}{9}} \nonumber\\
&=&\frac{32T}{9\sigma}\sum_{n}\int d|{\bf p}|{\bf p}^{2}\frac{1}{\log[(\pi^{2}T^{2}+\omega_{n}^{2}+{\bf p}^{2}+p_{R}^{2})/\Lambda^{2}_{QCD}]} \nonumber\\
&&\times\frac{1}{(\omega_{n}-\pi T)^{2}+{\bf p}^{2}}\frac{\Delta(p)}{\Delta^{2}(p)-\overline{p}^{2}}.
\end{eqnarray}
Since the left hand side is independent of $\sigma$, it is rather easy to seek the values of $\sigma$ satisfying this equation.

The parameters of our model are $\Lambda_{QCD}=738$ MeV and $t_R\equiv\log(p^{2}_{R}/\Lambda_{QCD}^{2})=0.1$. These values are the same as those used in Ref.\cite{KMT98}, where the trial mass function is also the same as ours.

The above equation has three types of solutions in accordance with the values of the temperature $T$ and the chemical potential $\mu$:\\
~~(1)~~~$\sigma=0$ \\
~~(2)~~~$\sigma=0~,~\sigma_{1}$ ~~~~~~~~~~~($0<\sigma_{1}$) \\
~~(3)~~~$\sigma=0~,~\sigma_0~,~\sigma_1$~~~~~~($0<\sigma_0<\sigma_1$) \\
In the case (1), $\sigma=0$ is always a minimum and the thermodynamical potential $\Omega$ is monotone increasing with $\sigma$. In the case (2), $\sigma=0$ is a maximum and $\sigma_1$ a minimum so that $\Omega$ is decreasing ($\sigma<\sigma_1$) and increasing ($\sigma>\sigma_1$). In the last case, $\sigma=0$ and $\sigma_1$ are minima and $\sigma_0$ is a maximum: $\Omega$ is increasing ($\sigma<\sigma_0$ and $\sigma>\sigma_1$) and decreasing ($\sigma_0<\sigma<\sigma_1$). Moreover the case (3) is discriminated between the case of $\Omega(0)>\Omega(\sigma_1)$ $(3^{-})$ and that of $\Omega(0)<\Omega(\sigma_1)$ ($(3^{+})$). 

Now we consider behavior of the thermodynamical potential when the temperature increases (the chemical potential is fixed). The numerical result shows three patterns:\\ 
~~(a)~~$(2)\to (1)$ or $(3^{-})\to (1)$,~~~($\mu<\mu_P$) \\
~~(b)~~$(3^{-})\to (3^{+})\to (1)$,~~~~~~~~~($\mu_P<\mu<\mu_0$) \\
~~(c)~~$(3^{+})\to (1)$ or $(1)$.~~~~~~~~~~~~($\mu_0<\mu$) \\
In the pattern (a), the order parameter $\sigma$ changes from $\sigma_1$ to $0$ at $T=T_c$ continuously so that the transition is of the second-order. On the other hand in (b), $\sigma$ changes from $\sigma_1$ to $0$ at $T=T_c$ discontinuously. This abrupt jump occurs under the condition: $\Omega(T_{c},\mu,\sigma_1)=\Omega(T_{c},\mu,0)$. This equation determines the critical temperature $T_c$ and this transition is of first-order. The last pattern does not give any phase transitions and corresponds to the supercooling. The $\mu_P$ in the above equations is a critical point between such two transitions and the point $(\mu_P,T_P)$ is called a tricritical point. 

\begin{figure}
 \includegraphics[width=\linewidth]{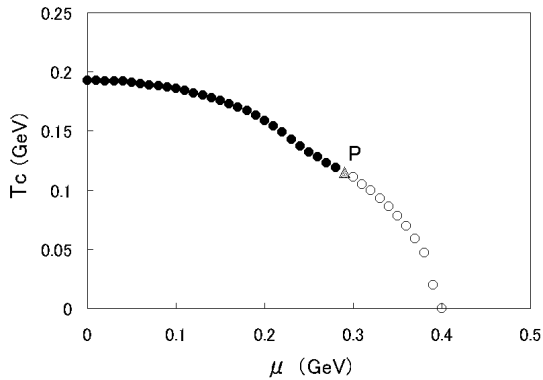}
 \caption{\label{fig:epsart}The schematic view of the phase diagram on the $T-\mu$ plane based on our numerical results. The point P denotes the tricritical one.}
\end{figure}
Numerical results are shown in Fig. 1, which is a phase diagram for the quark matter. The closed circles mean the second-order phase transition and the open ones the first-order one. The point "P" stands for the tricritical point. We have a second-order phase transition at $T_c=200$ MeV with $\mu=0$ and a first-order phase transition at $\mu_c=400$ MeV with $T=0$. The position of the tricritical point "P" is at ($T_P,\mu_P$)=(105 MeV, 300 MeV). These numerical results are consistent with those by the previous authors.

\section{LATENT HEAT}

In this section, we will calculate the latent heat generated across the first-order phase transition. For this purpose, we need to know the entropy density $S$ and the particle number density $\rho$ first. We note the following relation:
\begin{eqnarray}
d\Omega=-SdT-\rho d\mu.
\end{eqnarray}
Accordingly, we can calculate $S$ and $\rho$ from the thermodynamical potential as follows,
\begin{eqnarray}
S=-\frac{\partial\Omega}{\partial T},~~~~~~\rho=-\frac{\partial\Omega}{\partial \mu}.
\end{eqnarray}
Let us now introduce the difference between the thermodynamical potential in the symmetry broken phase and that in the symmetry one, 
\begin{eqnarray}
\Delta\Omega(T_c,\mu)\equiv\Omega(T_c,\mu,\sigma_1)-\Omega(T_c,\mu,0).
\end{eqnarray}
Similarly, we define the difference between the entropy density in the broken phase and that in symmetry one: $\Delta S(T_{c},\mu)=S(T_c,\mu,\sigma_1)-S(T_c,\mu,0)$.

We can calculate $\Delta S(T_{c},\mu)$ from the partial derivative of $\Delta\Omega$ with respect to the temperature $T$, 
\begin{eqnarray}
\Delta S(T_{c},\mu)=-\left( \frac{\partial \Delta\Omega}{\partial T}\right)_{\mu}(T_c,\mu).
\end{eqnarray}
Now the latent heat can be derived from this entropy difference as follows:
\begin{eqnarray}
Q_{l}=-T_{c}\Delta S(T_{c},\mu).
\end{eqnarray}

For example, we have drawn the $\Delta \Omega(T,\mu)$ as a function of the temperature with $\mu=0.35$ GeV in Fig.2.  Then we can evaluate the latent heat from the slope of the curve at the critical point $\Delta \Omega(T_c,\mu)=0$. The result of the latent heat per particle thus obtained is shown in Fig.3 as a function of the chemical potential $\mu$.
 \begin{figure}
 \includegraphics[width=\linewidth]{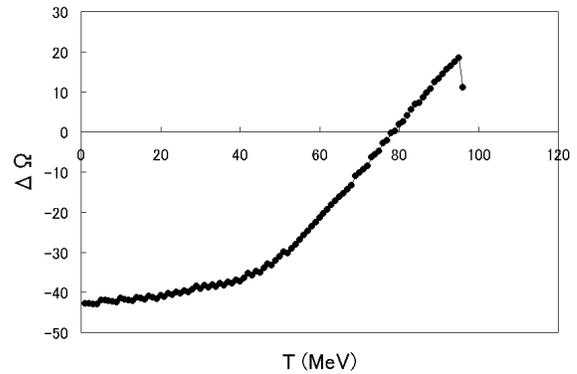}
 \caption{\label{fig:epsart}Temperature dependence of $\Delta\Omega$ at $\mu=350$ MeV. This is a case of the 1st-order phase transition. The curve passes through the horizontal line at the critical point $T_c$ and the slope of the tangential line gives the latent heat by Eq.(30).}
 \end{figure}
 \begin{figure}
 \includegraphics[width=\linewidth]{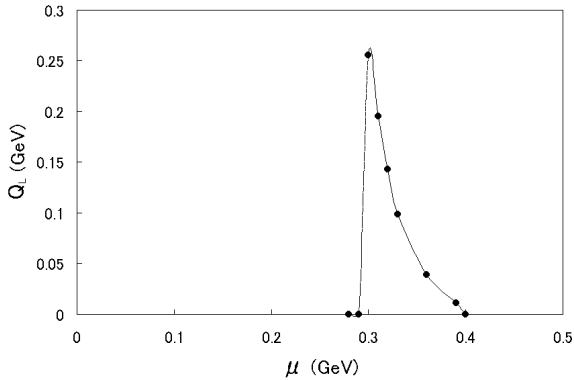}
 \caption{\label{fig:epsart}The latent heat $Q_l$ as a function of the chemical potential $\mu$.}
 \end{figure}
It tells us that the latent heat is not given off in the second-order phase transition ($\mu<\mu_P=0.30$ GeV). The latent heat emerges abruptly with large value ($\sim 250$ MeV) near the tricritical point ($\mu_P=0.30$ GeV). When the chemical potential is increased, it decreases and vanishes at $\mu=0.40$ GeV. In order to realize this behavior, let us consider the following relation,
\begin{equation}
Q_{l}=T_{c}\frac{\partial \Delta\Omega(T_{c},\mu)}{\partial T_c},
\end{equation}
which is derived from Eqs.(28) and (29). When $\mu$ is increased in the region of the first order transition, the figure of $\Delta\Omega(T)$ becomes shallow as a function of $T$ (see Fig.2): This means that the derivative ${\partial \Delta\Omega(T_c)}/{\partial T_c}$ decreases with increasing $\mu$. Therefore the latent heat decreases with increasing $\mu$. Our numerical result shows that the latent heat has a singularity near the tricritical point and plays an important role to identify the tricritical point.

\section{SUMMARY and DISCUSSIONS}

We have studied the latent heat in the chiral phase transition in this paper. We have taken the QCD-like gauge field theory as an effective theory and have performed the numerical calculations using the mean field (WKB) approximation. We have calculated the latent heat generated in the first-order phase transition. The latent heat is enhanced near the tricritical point and decreases as the chemical potential increases. However, it should be pointed out that more accurate numerical simulations will be necessary for the detailed behavior of the latent heat near the tricritical point.

We must now note the existence of the color superconducting state which has been disregarded in our numerical calculations \cite{BL84}-\cite{ARW98}. The color superconducting state is considered to be able to exist in the region at low temperature and high density apart from the tricritical point. Hence, we can not apply our results directly on the actual physical phenomenon, in which color superconductivity is involved. However, the phase transition on the color superconductivity is of the second-order so that it does not give off the latent heat originated on this transition.

It is pointed out that the values of T and $\mu$ realized in high energy heavy-ion collisions may be close to the tricritical point \cite{SRS98}-\cite{HJSSV98}. Therefore, it may be possible to observe some signals originated in the generation of the latent heat because of the fact that the latent heat becomes large near the tricritical point.

\begin{acknowledgments}
We would like to thank Nuclear Theory Group at Kochi University for helpful discussions. One of us (H.K.) is also grateful to Mr.H.Akaike and Mr.T.Hama for kind help with his computer work.
\end{acknowledgments}

\end{document}